\documentclass{iau}

\usepackage{graphicx,natbib}

\newcommand{\apj}{ApJ}           % Astrophysical Journal
\newcommand{\apjl}{ApJ}           % Astrophysical Journal
\newcommand{\mnras}{MNRAS}       % Monthly Notices of the RAS

\newcommand{\aj}{AJ}

\newcommand{\apjs}{ApJS}           % Astrophysical Journal

\newcommand{\AD}{\mbox{ATLAS$^{\rm 3D}$}} 
\newcommand{\Reff}{\mbox{R$_{\rm e}$}}

\def\sbr{${\rm mag\,\,arcsec^{-2 }}$\ }
\newcommand{\Mo}{\mbox{M$_{\odot}$}}

\title{Applying galactic archeology  to massive galaxies using deep imaging surveys}

\author[Duc]{Pierre-Alain Duc$^1$}

\affiliation{$^1$AIM Paris-Saclay, CEA:Irfu, CNRS/INSU, Universit\'e Paris Diderot, France. email: {\tt paduc@cea.fr}}

\pubyear{2014}
\volume{311}
\jname{Galaxy Masses as Constraints of Formation Models}
\editors{M. Cappellari \& S. Courteau, eds.}

\begin{document}

\maketitle

\begin{abstract}
Various   programs aimed at exploring the still largely unknown low surface brightness Universe with deep imaging optical surveys have recently  started. They open a new window  for studies of galaxy evolution,  pushing the technique of galactic archeology outside the Local Group (LG). 
The method, based on the detection and analysis of the diffuse light emitted by collisional debris or extended stellar halos (rather than on stellar counts as done for LG systems),  faces however a number of technical difficulties, like the contamination of the images by reflection halos and Galactic cirrus. I review here the on-going efforts to address them and  highlight the preliminary promising results obtained with a systematic survey with MegaCam on the CFHT of nearby massive early-type galaxies done as part of the \AD, NGVS  and MATLAS collaborations.

\end{abstract}

%\firstsection
\section{Introduction}
The processes responsible for galaxy evolution, their formation, growth and death,  may be be studied either probing slices of the Universe at different epochs through redshift surveys or scrutinizing and dissecting  todays galaxies to infer their past mass assembly. The success of the latter method strongly depends on our ability to detect   structures   located in the very outskirts of galaxies: the extended stellar halos and   collisional debris that keep the memory of past mergers. Their number and shape inform us on the mechanisms driving the growth of galaxies: wet vs dry, minor vs major mergers, secular processes, etc. 
Unfortunately, such structures are of low surface brightness (LSB), and are difficult to detect unless their are resolved into individual stars, like for Local Group systems. In the last decades, thanks in particular to the great image quality of the Hubble Space Telescope  and dedicated  ground-based surveys, galactic archeology made tremendous progress, and the Gaia space mission  will boost it even more. However the method is confined to galaxies located within a few Mpc from the Milky Way. The most massive ones, in particular lenticulars and ellipticals the origin of which remains actively debated, are located too far away to be mapped with stellar counts. 
Until Extremely Large Telescopes become available,  the only way to detect their outer structures is through their diffuse light. 
 Fortunately,  the  availability of large field of view optical cameras, and the development of  clever observing and data reduction techniques, have led to a gain of several magnitudes in surface brightness limits with respect to classical imaging and now allow us to detect this diffuse light. 
 %I review here these new imaging efforts, giving an emphasis on initial results obtained with surveys made with the MegaCam camera on the CFHT that focus on massive early-type galaxies. 

\section{On-going deep imaging surveys}

\begin{figure}[h]
\centering
\includegraphics[width=0.93\columnwidth]{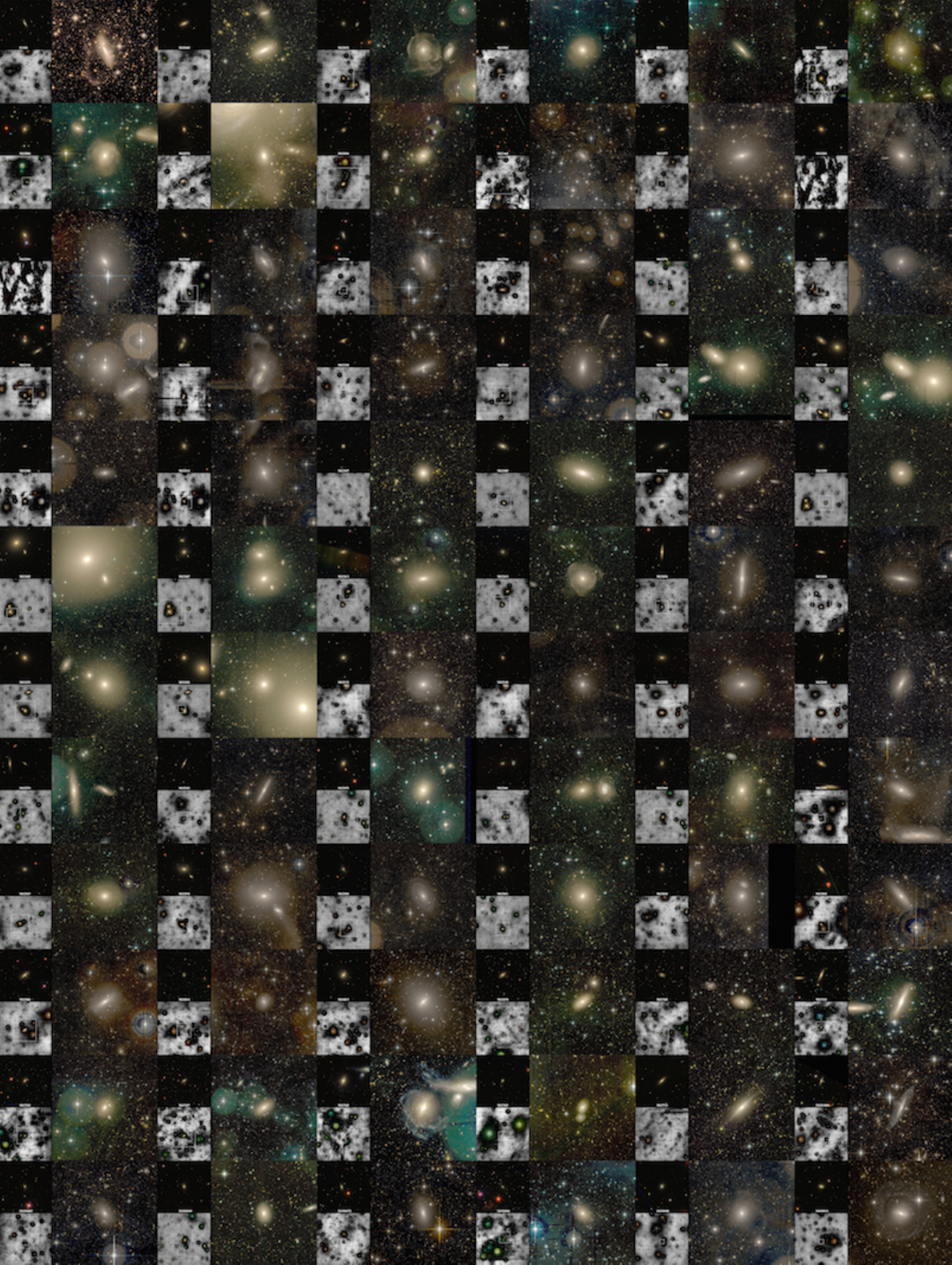}
\caption{A catalog of early-type galaxies from the \AD\ survey. Each sub-panel displays a galaxy as seen by  the SDSS  (top-right) and the MegaCam/CFHT camera   (right). A grey scale g-band MegaCam image of the field around the ETG is shown to the bottom-left. A high resolution version of this poster may be obtained at  http://irfu.cea.fr/Projets/matlas/MATLAS/Outreach/Outreach.html}\label{fig1}
\end{figure}

\subsection{Achievements: an increased sensitivity to LSB  structures }

Thanks to its large coverture of the sky, homogeneity and user friendly access to its archives, the Sloan Digital Sky Survey \citep[SDSS,][]{York00}, has become {\it the} unavoidable resource  for  optical studies of galaxies in the nearby Universe. However, its sensitivity to extended low surface brightness features is very limited, largely hampering their direct use  for galaxy archeology. 
A number of on-going projects aim at getting much deeper images. They make use of stacking techniques \citep[e.g.][]{Kim13}, refurbished or dedicated instruments \citep[e.g.][]{Mihos05,vanDokkum14},  extremely long exposures on small telescopes  \citep[e.g.][]{Martinez-Delgado10}, or LSB-optimized surveys, such as the Next Generation Virgo Cluster Survey \citep[NGVS, ][]{Ferrarese12}, or the MegaCam component  of the  \AD\ project \citep{Duc14b}, to be completed thanks to the MATLAS CFHT Large Programme.

Fig.~\ref{fig1} illustrates the gain of the LSB-optimized MegaCam survey, which reaches about 28.5--29~\sbr\ in the g--band  to be compared to about 26.5  for the SDSS.

One should note  that the  sensitivity to LSB structures does not only depend on the telescope size and exposure time, but is largely driven by the handling of systematics in the observations. Consequently, the limiting surface brightness   is particularly difficult to estimate, partly  explaining the large range of published values in the literature \citep[see][ and references therein]{Duc14b}.
Whether deep ground-based  imaging surveys   can reach the limiting surface brightness obtained in Local Galaxies with stellar counts, i.e. 32~\sbr, as claimed for instance for the amazing Dragonfly experiment (Abraham et al., this volume), still needs to be clarified.  

\subsection{Challenges: halos and cirrus}

In addition to the classical issues of large-scale variations of the background (not easily) addressed by various flat--fielding or sky subtraction techniques,  deep imaging is hampered by two  stumbling blocks that considerably limit their capability of detecting diffuse extended structures: artificial  halos due to internal reflections in the instrument and Galactic cirrus emission.

\begin{figure}
\centering
\includegraphics[width=\columnwidth]{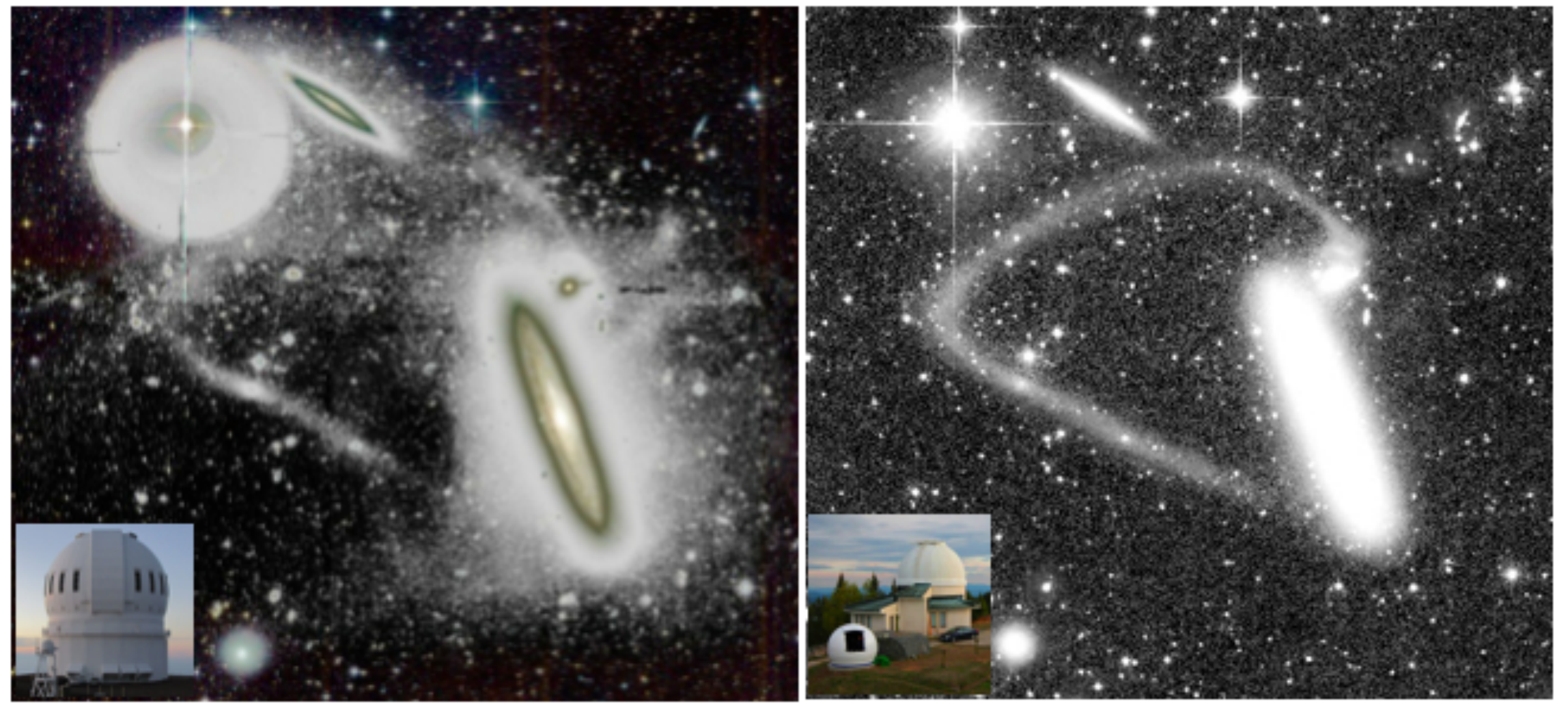} 
\caption{Comparison between the image of the spiral galaxy NGC~4216 obtained  with MegaCam at the CFHT as part of the NGVS (Paudel et al., 2013, left, 1 hour exposure time) and at the amateur Irida observatory in Bulgaria (courtesy V. Popov and E. Ivanov, right, 20 hours exposure time). The latter image is not as deep as the first one, but looks cleaner due to the absence of the prominent reflection halos that plague the CFHT image. }\label{fig2}
\end{figure}

The light of bright stars but also the nucleus of galaxies is reflected between the CCD and the different optical elements of the camera, causing the presence on deep images, such as the ones produced by MegaCam,  of multiple more or less concentric halos, with sizes reaching several arcmin. 
As  illustrated in Fig.~\ref{fig2}, they may hide fine structures like tidal tails or be mistaken with the  stellar halos that surround galaxies, or the intra-cluster light in which cluster galaxies are  embedded.  A special coating of CCDs and filters or a completely different optical construction, as done for the Dragonfly  telephoto lenses, may lower the impact of such instrumental signatures. Besides, there are on-going efforts to  a posteriori remove them  thanks to a  modeling of the instrument and ray tracing experiments.

Because it is unavoidable and extremely difficult to subtract,  an even more serious issue  is the pollution by Galactic cirrus. Dust clouds around our own Milky Way scatter light in the optical regime. They show up on deep images as complex, extended, multicolored, filamentary structures that may be mistaken with stellar tidal tails
\citep[see examples in ][ and in Fig.~1]{Duc14b}.
 The Planck survey revealed that Galactic cirrus is present over a large fraction of the sky.   Cirrus emission may thus potentially be an issue for all  deep imaging surveys. 
In that respect, it is somehow surprising or reassuring that the    extremely deep image of M101 obtained with Dragonfly \citep{vanDokkum14} does not show clear evidence  of cirrus emission. 

\section{Results}

Despite the technical  issues mentioned above, the asset of deep imaging of galaxies is undeniable. It reveals a wealth of so far unknown structures that inform us about the various events that build galaxies, but also raises issues on how galaxies are classified.

\subsection{Is traditional morphological classification  challenged by deep imaging?}

\begin{figure}
\centering
\includegraphics[width=\columnwidth]{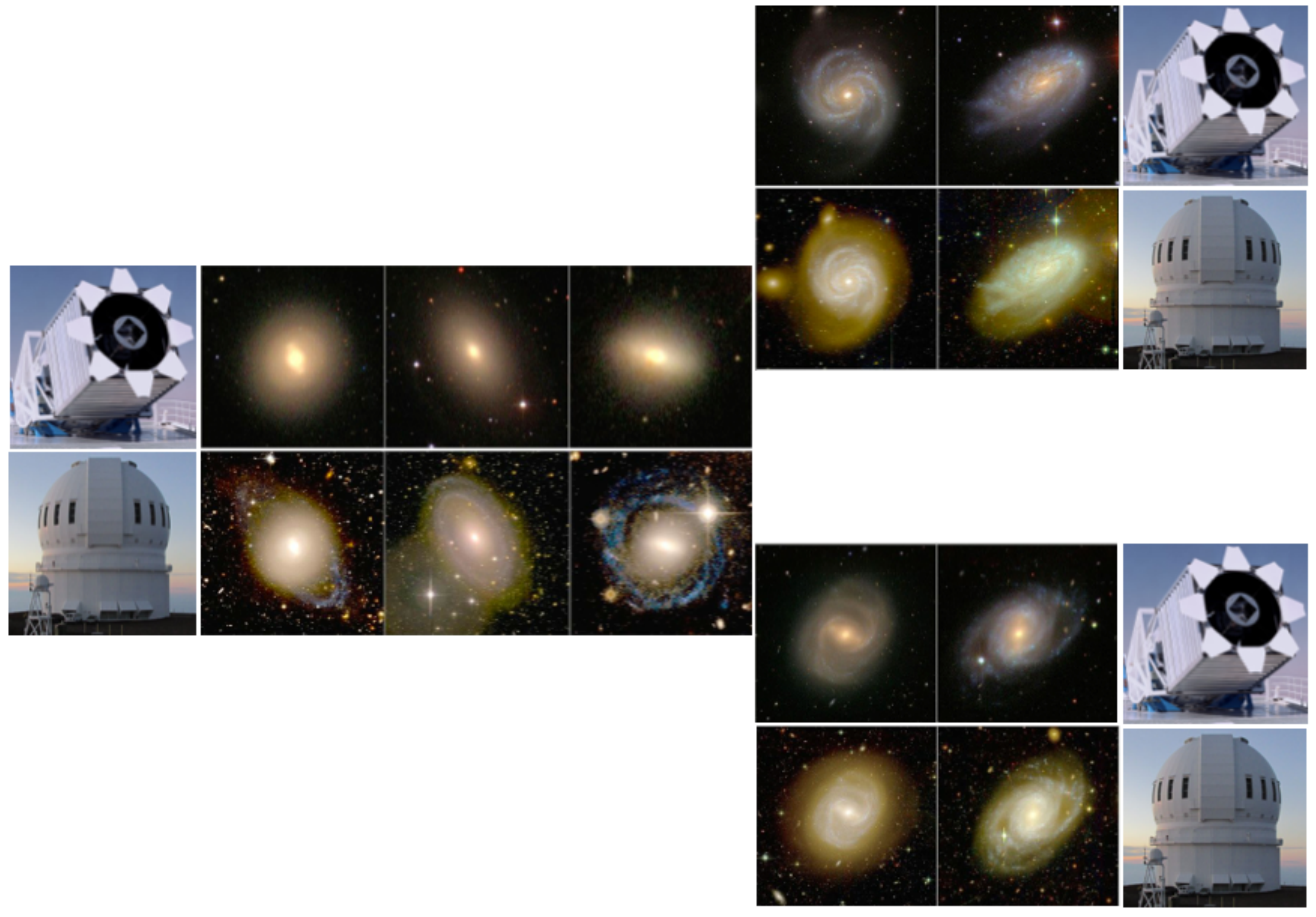} 
\caption{The Hubble diagram illustrated with  images of 7 early and late type galaxies obtained with the Sloan (SDSS, top panels) and MegaCam at the CFHT (bottom panels, same objects as above but with slightly larger fields of view). For the specific galaxies shown here, {\it the red and dead} ETGs turn out to exhibit a blue, star--forming,  component on the  the CFHT images, while the blue, star-forming, spirals appear to be embedded in a red stellar halo.} \label{fig3}
\end{figure}

The  classification of galaxies traditionally relies on morphological criteria (bulge to disk ratio, bar strength) that are checked inspecting optical images. However, the detectability of the discriminating structures depends on the depth of the available images. Indeed, as shown on Fig.~\ref{fig3}, some apparently red galaxies classified as early-type by a visual inspection of SDSS images exhibit on the deeper MegaCam images so-far undetected low surface brightness blue spiral-like structures. This does not imply that morphological classification is unfounded, but obliges us to take into consideration on which data set it was done.  

\subsection{Are scaling relations affected by deep imaging?}

\begin{figure}
\centering
\includegraphics[width=\columnwidth]{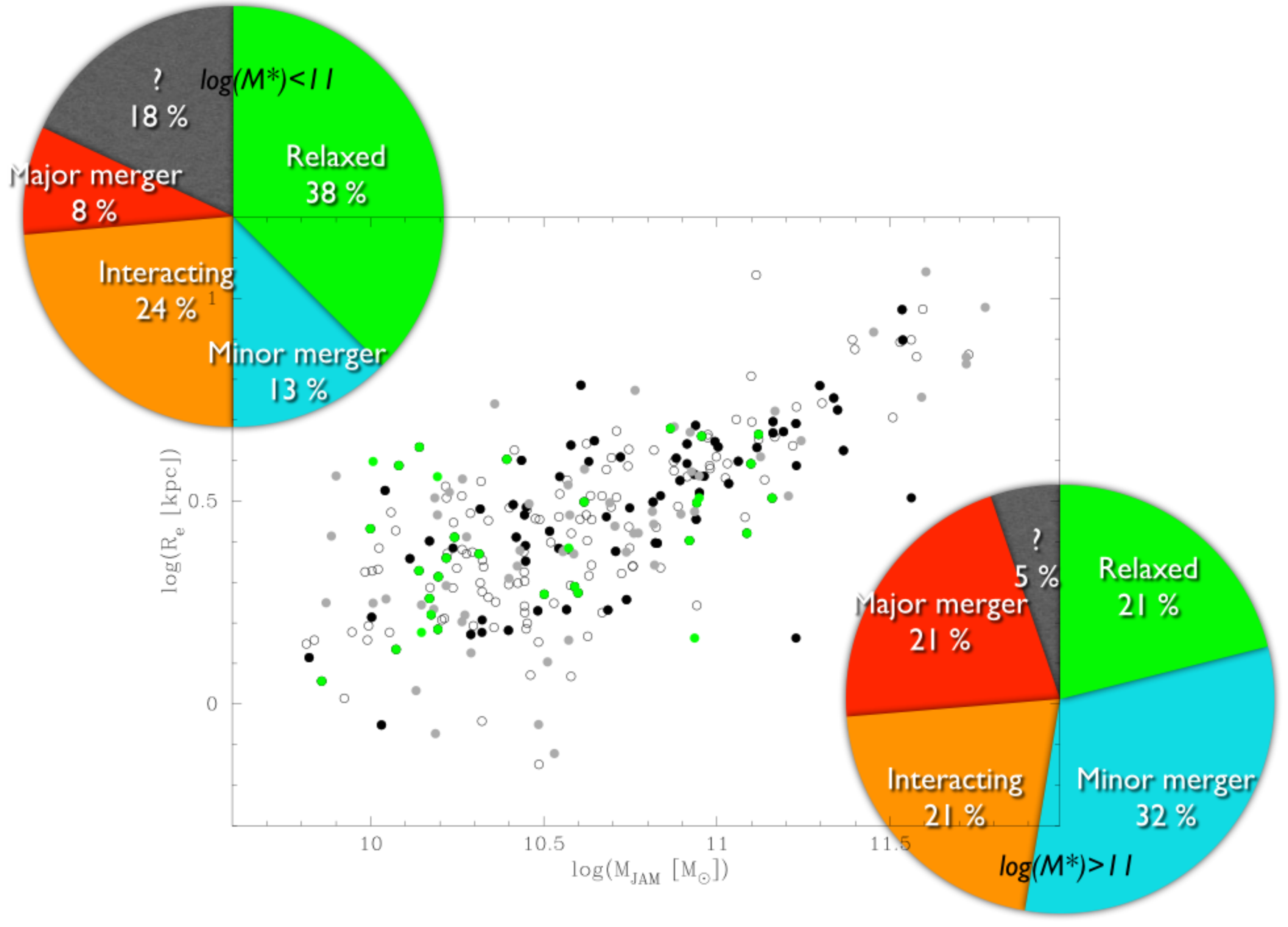} 
\caption{Dynamical mass, M$_{\rm JAM}$,  versus  the effective radius, \Reff\  \citep[see][]{Cappellari13}, of nearby early-type galaxies.    Open circles correspond to the full \AD\ sample, and filled discs are galaxies with available deep imaging. Relaxed galaxies showing no sign of tidal perturbations even with the deep imaging are shown in green. The cake diagrams on each side of the plot display the statistical results of the galaxy classification made with the deep images. }\label{fig4}
\end{figure}

Stellar mass and  sizes are other fundamental parameters used in categorizing  galaxies and/or   determining universal scaling relations that blur their differences (see example of the mass--size relation of ETGs in Fig.~\ref{fig4}). The values of such parameters depend again in principle on the depth of the images used to determine them. 
We quantified this measuring the  diffuse light component beyond an  isophote  of 26~\sbr, i.e. the component not detected in traditional shallow imaging surveys. The extra light ranges between 5 and  20 percent (in the g--band), depending on the galaxy mass. It is  higher in dense environments such as the Virgo cluster, but there  the light of the very extended outer halo mixes with the intra-cluster light, and its importance is difficult to estimate.   

For galaxies with masses below $10^{11}~\Mo$, we do not see significant differences between the effective radius of the \AD\ ETGs determined with MegaCam and that published the literature. However a significant systematic excess of \Reff\ of up to a factor of 1.7 is found above this threshold \citep{Duc14b}. 

Such updates of the mass and size  do not significantly alter the global scaling relations given their intrinsic scatter. However, the variations of the mass/size excess as a function of  mass, morphological type, etc., may give clues on how galaxies were assembled. 

\subsection{Are fine structures present around all types of galaxies?}

Besides its depth, the originality of the   MegaCam \AD\ survey  lies in the high number of observed galaxies: a complete sample of  260   nearby ETGs (plus about 100 spirals located in their field) when it  is completed. This enables  to  get reliable statistical properties, to be compared with predictions  from cosmological models  of galaxy formation and associated numerical simulations. 
The latter predict that most, if not all, massive galaxies should be surrounded by stellar streams, i.e. the remnants  of the multiple  (mostly minor)  mergers that presumably build them. The  sensitivity limit of currently available surveys, including those presented here, and the fading of tidal debris with time   allow us to  detect the brightest streams, that trace relatively recent collisions (that occurred at redshift below 0.5--1). 

Based on the catalog of 92 galaxies presented in \cite{Duc14b}, we computed  the fraction of galaxies showing signs of tidal perturbations. Galaxies were classified as on-going interacting systems, major mergers, minor mergers, and fully relaxed systems depending on the shape and number of fine structures around them. Preliminary results for two mass ranges are shown on Fig.~\ref{fig4}. The  galaxies in the high mass bin appear to have experienced more recent minor and major mergers. 
The fraction of fully relaxed systems drops by a factor of two from the low to the high mass bin.  
Similar trends are found going from fast to slow rotators, gas--poor to gas--rich objects, though selection criteria might still bias these results.
 Getting reliable statistics requires the completion of the survey. 

%\section{Conclusions}

\section*{Acknowledgements}
\noindent
I am grateful to all my collaborators on this program made within the  framework of the \AD, NGVS and MATLAS projects. Many thanks in particular to Jean-Charles Cuillandre and Emin Karabal.

%\bibliographystyle{mn2e}
%\bibliography{../Biblio/all}

\end{document}